# Ultralow-power cryogenic thermometry based on optical-transition broadening of a two-level system in diamond

Yongliang Chen,[1] Simon White,[1,*] Evgeny A. Ekimov,[2,3] Carlo Bradac,[4] Milos Toth,[1,5] Igor Aharonovich[1,5] and Toan Trong Tran[1,6,*]

1. School of Mathematical and Physical Sciences, University of Technology Sydney, Ultimo, NSW, 2007, Australia

2. Institute for High Pressure Physics, Russian Academy of Sciences, Troitsk 108840, Moscow, Russia

3. Lebedev Physics Institute, Russian Academy of Sciences, Moscow 117924, Russia

4. Trent University, Department of Physics & Astronomy, 1600 West Bank Drive, Peterborough, ON, K9L 0G2

5. ARC Center of Excellence for Transformative Meta-Optical Systems (TMOS), Faculty of Science, University of Technology Sydney, Ultimo, NSW, 2007, Australia

6. School of Electrical and Data Engineering, University of Technology Sydney, Ultimo, NSW, 2007, Australia

*Corresponding author: simon.white@uts.edu.au

*Corresponding author: trongtoan.tran@uts.edu.au

**Abstract**

Cryogenic temperatures are the prerequisite for many advanced scientific applications and technologies. The accurate determination of temperature in this range and at the sub-micrometer scale is, however, non-trivial. This is due to the fact that temperature reading in cryogenic conditions can be inaccurate due to optically-induced heating. Here, we present an ultralow-power, optical thermometry technique that operates at cryogenic temperatures. The technique exploits the temperature-dependent linewidth broadening measured by resonant photoluminescence of a two-level system—a germanium-vacancy color center in a nanodiamond host. The proposed technique achieves a relative sensitivity of ~20% $K^{-1}$, at 5 K. This is higher than any other all-optical nanothermometry method. Additionally, it achieves such sensitivities while employing excitation powers of just a few tens of nanowatts—several orders of magnitude lower than other traditional optical thermometry protocols. To showcase the performance of the method, we demonstrate its ability to accurately read out local differences in temperatures at various target locations of a custom-made microcircuit. Our work is a definite step towards the advancement of nanoscale optical thermometry at cryogenic temperatures.

**Introduction**

Cryogenic temperatures—ranging from absolute zero to 120 K—present a crucial window for the exploration of cutting-edge applications, both fundamental and practical, that rely on low temperature operating conditions.[1-3] Advanced realizations, for example in optoelectronics, include the recent discovery of unconventional superconductivity in twisted bilayer graphene,[4] the observation of quantum Hall effect in graphene [5] and the demonstration of atomically-thin quantum light-emitting diodes[6] to name a few. To facilitate the rapid advancement of this field, appropriate nanoscale thermometers are required to monitor and determine the local temperatures of these devices with accuracy. To this end, contact-based thermometers such as micro-thermocouples, microresistor probes or scanning thermal microscopes have traditionally been employed [REF].[7-9] These contact-based thermometers, however, display complex probe-sample heat transfer mechanisms that are difficult to model or account for as the samples reach the sub-micrometer scale. Optical thermometry eliminates this problem by employing fluorescent nanoprobes of negligible thermal load as the sensing elements. Among optical nanothermometers, those based on photoluminescent nitrogen-vacancy (NV) and group IV color centers in diamond stand out due to a series of desirable properties.[10] These defects can emit light at different wavelengths and display temperature-dependent spin and spectral properties that enables parallel, multicolor imaging and temperature monitoring. They are also photostable over several hundreds of K and the diamond nanohosts are nontoxic and easy to functionalize to specific environments or moieties. However, they are not free of limitations. Nanothermometers based on diamond NV and group IV color centers display excellent sensitivity at room temperature or above, but their performance drops for temperatures below ~100 K due to a significant reduction in sensitivity.[11, 12]

Recently, a range of fluorescent thermometers including lanthanide phosphonate dimer,[13] metal-organic frameworks (MOFs),[14-18] and doped metal oxide phosphors[19] have been developed to measure temperatures in cryogenic conditions. While some of these thermometers achieve sensitivities as high as ~32% $K^{-1}$,[14, 16] they typically are sub-millimeter in size—which precludes their use at the nanoscale. Here, we propose a cryogenic thermometer based on the photoluminescence linewidth broadening of a single negatively-charged germanium vacancy (GeV) color center in nanodiamonds, with an excellent sensitivity of ~20% $K^{-1}$. This is achieved through resonant excitation of the center using ultralow laser powers (~tens of nanowatt)—several orders of magnitudes lower than other competitive methods. Additionally, the technique can reach nanoscale spatial resolutions, since the active sensing elements are single, atom-like GeV centers that could potentially be hosted in nanodiamonds just a few nanometers in size.[20]

**Result and discussion**

The nanothermometers we propose consist of GeV color centers hosted in nanodiamonds that were grown via high pressure high temperature (HPHT, c.f. Methods). Briefly, a powder mixture of adamantane ($C_{10}H_{16}$) and tetraphenylgermane ($C_{24}H_{20}Ge$) was pressed into a pellet and placed inside a titanium capsule. The capsule was then compressed at high temperature (1800–2000 °C) and high pressure (9 GPa) inside a reactor and cooled under high pressure to ambient temperature. The resulting nanodiamonds were then dispersed in isopropanol alcohol (IPA), drop-casted on a clean silicon substrate and left to dry on a hotplate at 60 °C to completely remove the residual solvent.[21] **Figure S1a** in the Supporting information shows the sizes of the nanodiamonds ranging from one hundred to a few hundreds of nanometres. A representative Raman spectrum (**Figure S1b**) collected from a randomly chosen nanodiamond reveals the sharp diamond characteristic peak at ~ 1329 cm$^{-1}$, indicating that the nanodiamonds are highly crystalline.[21]

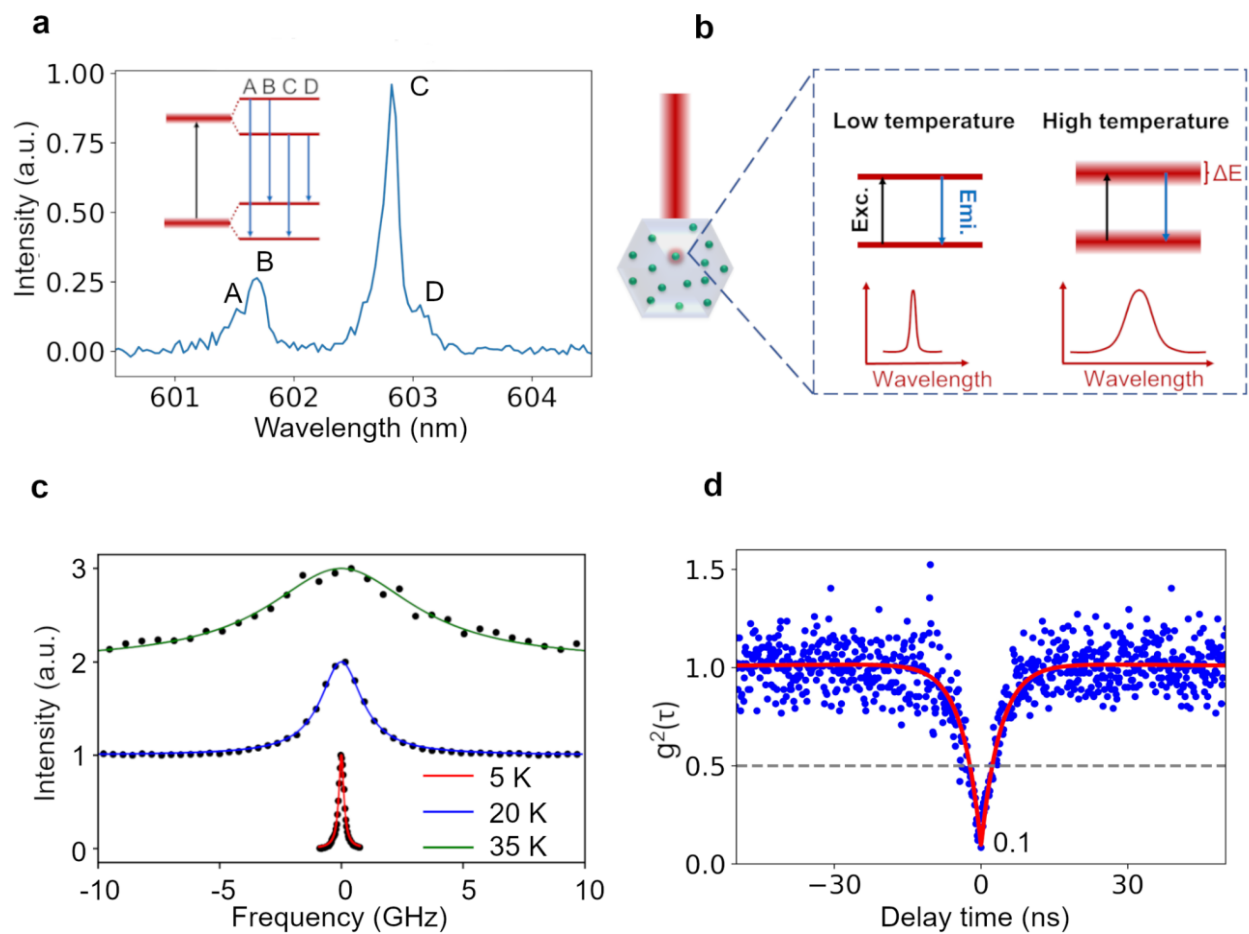

*Figure 1. Emission from a single GeV center in a nanodiamond at cryogenic temperatures. a) A representative spectrum acquired by a spectrometer with a grating of 1800 g/mm under the off-resonance excitation of a 532-nm, 300-μW laser at 5 K. Inset: schematic depicting the manifold level system of a GeV color center. b) linewidth broadening due to phonon-mediated coupling. c) Second-order autocorrelation measurement from resonant excitation of the GeV center under 200 nW of laser power. The dip of 0.1 indicates the single-photon nature of the emission. d) Two representative PLE spectra of transition C were acquired by a 15-nW laser*

*scanning with a step size of 16.5 MHz at 5 K (blue line) and 24.8 MHz at 35 K, respectively (green line), displaying obvious linewidth broadening of the transition C emission with the temperature.*

The negatively-charged GeV belongs to the group IV diamond color centers alongside the silicon-vacancy (SiV), tin-vacancy (SnV) and lead-vacancy (PbV) center. These defects consist of an interstitial group IV atom (Si, Ge, Sn or Pb) located within two adjacent missing carbons, in the split-vacancy configuration of $D_{3d}$ symmetry.[22] A representative fluorescence spectrum of an ensemble of GeV color centers taken at room-temperature is shown in **Figure S2**. The spectrum clearly features a sharp zero-phonon line (ZPL) alongside a relatively weak phonon sideband (PSB)**.** Owing to its inversion symmetry, the dipole moment of the GeV color center is unsusceptible to local stray fields of the first order created, e.g., by trapped charges and impurities in the diamond lattice.[22] As a result, its ZPL is spectrally stable, making this color center an ideal choice for our demonstration. Furthermore, the ZPL accounts for ~70% of the total fluorescence of the center,[23] which is desirable for several applications in quantum optics and nanophotonics.[24-27] To characterize the optical response of the GeV centers, we loaded the sample into a commercial closed-cycle helium gas cryo-refrigerator (c.f. Methods) with an optically-accessible window. We used a high numerical aperture objective (0.82 NA) to excite individual nanodiamonds and collect their fluorescence at 5–35 K. To showcase the versatility of our technique, we chose nanodiamonds with ensembles of GeVs rather than single ones. This is because our method is inherently selective and allows for the optical interrogation of a single emitter at a time (see below). **Figure 1a** shows a representative spectrum of an ensemble of GeV centers at 5 K, excited with a 532-nm continuous-wave (CW) laser. The four distinctive peaks indicate the four well-known optical transitions of the GeV center at ~601–603 nm, which are labelled A, B, C and D (**Figure 1a**, inset).[26] Note that in the measurement, each peak shows inhomogeneous broadening due to the presence of several GeV centers in the same nanodiamond.[24] The broadening is however too small to be discriminated by our spectrometer equipped with a 1800 g/mm grating (~0.035 nm resolution).

To spectrally isolate an individual GeV from the other centers in the ensemble, we employ resonant photoluminescence excitation (PLE), which is widely used in quantum optics to interrogate a single atom/molecule at a time.[28, 29] In the nanodiamond, the presence of local strain and electric fields causes the ZPL of each emitter to shift from that of any other center by up to tens of gigahertz. This relatively large energy separation and the absence of spectral diffusion makes it possible to isolate a single GeV's optical transition from any other and monitor it as temperature changes. In this study, we employ a narrow laser with a linewidth of ~200 KHz to scan across the ZPL of a single GeV optical transition and measure its spectral linewidth. The spectral broadening from this transition is determined mainly by two factors, i) the linewidth defined by the Fourier transform of the emitter's lifetime, and ii) the linewidth originating from the pure dephasing of the emitter, due to phonons.[30] The former factor does not show any significant temperature dependence up to 300 K and is treated as a constant in our fit (see discussion below).[25] The latter, however, is strongly affected by temperature variations and displays both linear and cubic temperature-dependent terms—the combination of which can be determined from the PLE scans.[31] **Figure 1b** summarizes the basic idea of our approach. At low temperatures, the spectral linewidth of a single GeV is relatively narrow; it however broadens significantly at higher temperatures. To demonstrate the change in linewidths as the temperature varies, we carried out the PLE measurements using spectral filtering to reject the reflection of the resonant excitation laser. Specifically, we sweep the frequency of the laser across the ZPL while we collect the fluorescence signals from the phonon sideband (PSB) of the emitter into a single-photon avalanche photodiode (SPADP). Since each emitter has four optical transitions (A, B, C and D in Fig. 1A), we focus on transition C, which is the brightest in intensity. **Figure 1c** shows the ZPL lineshape of transition C from a representative GeV color center taken at an excitation power of 15 nW and at various temperatures: 5 K (bottom), 20 K (middle) and 35 K (top). We emphasize that the excitation power we used in this measurement is the lowest across any of the existing all-optical nanothermometry techniques. The low excitation power is key to obtain an accurate readout of the temperature as it eliminates the issue of power-induced heating caused by laser excitation. The absence of spectral diffusion—which would result in the lineshape of the transition to be Gaussian—allows us to fit these peaks with single Lorentzian functions.[31] The measured linewidths are: 231 MHz at 5 K, 2081 MHz at 20 K, and 7380 MHz at 35 K—which corresponds to a 32-fold relative change over the temperature range 5–35 K. Such a strong temperature dependence makes monitoring the linewidth of a single GeV photoemission, via PLE, an excellent strategy for nanoscale thermometry. To confirm single-photon emission from the target GeV, we performed a second-order autocorrelation measurement. We set the laser excitation wavelength to be resonant to that of the center's ZPL and recorded the photoluminescence signals from the phononic sideband (PSB) of the emitter. Photon correlations were measured using two single-photon detectors in a Hanbury-Brown and Twiss interferometer connected to a high-resolution time tagger (cf. Methods). The strong photon antibunching dip of ~0.1 at zero-delay time in the autocorrelation measurement in **Figure 1d** confirms the emitter to be a single-photon source.

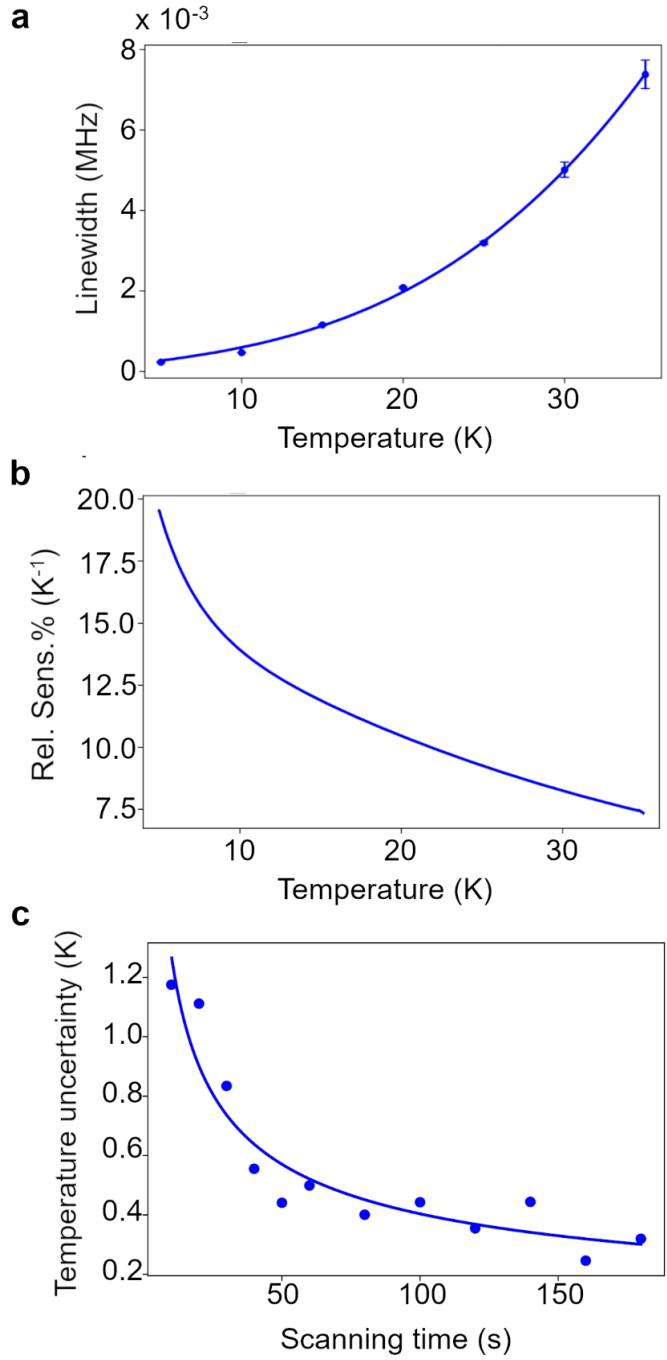

*Figure 2.* Characterization of the nanothermometer based on the linewidth broadening of transition C from a single GeV center. a) Temperature dependence of linewidth from the same single transition line in the nanodiamond. Each linewidth is extracted from the Lorentzian fit of the corresponding PLE spectrum. The PLE spectra are acquired with a scanning time of 180 s under a 15-nW laser. The blue line is fitted with the function $\gamma_{tot} = c + bT + aT^3$. b) Relative sensitivity of the thermometer as a function of temperature. c) Temperature uncertainty ($\sigma_T$) of the thermometer as a function of the scanning time ($t_m$) at 5 K. A shot noise of 4 K·Hz$^{-1/2}$ is extracted from the equation fit of $\eta_T = \sigma_T\sqrt{t_m}$.

We now turn our attention to the calibration of the nanothermometer. To establish the temperature response of the thermometer, we extract the linewidths from the PLE scans across the transition C of the emitter while increasing the temperature in steps of 5 K, from 5 to 35 K. **Figure 2a** shows that as the temperature rises, the linewidth of the ZPL increases non-linearly. As mentioned earlier, the temperature dependence of the linewidth is due to two factors and three main components: i) $\gamma_{lt}$, the constant linewidth associated to the temperature-invariant lifetime of the emitter, iia) $\gamma_{1st}$, a linear temperature-dependent contribution to the linewidth of the first-order, and iib) $\gamma_{2nd}$, a cubic temperature-dependent contribution to the linewidth due to the second-order electron–phonon interactions with E-symmetric phonon modes.[31] The fitting equation is thus: $\gamma_{tot} = \gamma_{lt} + \gamma_{1st} + \gamma_{2nd} = c + bT + aT^3$, where $\gamma_{tot}$ is the total linewidth, $c$ is a constant, while $b$ and $a$ are the coefficients of the second and third terms in the equation, respectively. To determine the constant $c$, we first extracted the lifetime of the emitter from the second-order autocorrelation function in **Figure 1d**, which gave a value of 3.9 ns. The constant $c$ was estimated to be 41 MHz given that: $c = \gamma_{lt} = \frac{1}{2\pi\tau}$, where $\tau$ is the emitter's lifetime. The experimental data was then fitted to the equation for $\gamma_{tot}$. The fit was used as the temperature calibration curve for the thermometer. **Figure 2b** shows a plot of the relative sensitivity as a function of temperature. The relative sensitivity is a key performance metric commonly used in nanothermometry and is defined as $S_r = \frac{1}{\gamma_{tot}} \frac{d\gamma_{tot}}{dT}$. The relative sensitivity of our technique is 20% K$^{-1}$ at 5 K—the highest reported value among all-optical nanothermometry methods. Note that while there are alternative techniques with higher relative sensitivity, ~31%[16] and ~32%[14], they operate at size regimes larger than the nanoscale.

Another important performance index of a thermometer is its temperature resolution. It must be noted that temperature resolution, $\eta_T$, unlike relative resolution, is a relative quantity that can be improved by various means. These include, e.g., increasing the excitation power and/or the integration time, improving the collection efficiency of the optical setup, etc. Specifically, the temperature resolution relates to the integration time $t_m$ and the standard resolution of the measurement, $\sigma_T$, via the following expression $\eta_T = \sigma_T \sqrt{t_m}$. By varying the laser scanning time in the PLE measurement and estimating the standard deviation for the linewidth $\gamma_{tot}$, we can extract the uncertainty of the temperature readout. **Figure 2c** shows this relation with a shot noise fit function $\frac{1}{\sqrt{t_m}}$, where $t_m$ is the laser scanning time. The fit is in good agreement with our experimental data, showing a temperature resolution $\eta_T$ =4 K·Hz$^{-1/2}$. This value is relatively modest compared to other all-optical nanothermometers. This follows from the fact that we only excite a single center for a relatively short time, rather than an ensemble of several emitters. However, and importantly, the excitation power we used in this study is only ~tens of nanowatt—several orders of magnitude lower than other techniques. We can objectively take both factors into consideration by estimating the temperature resolution relative to power. This is comparatively high for our technique, of ~6.2 K·Hz$^{-1/2}$·W·cm$^{-2}$, more than three orders of magnitude greater than that of the next best nanothermometer (1.4 × 10$^4$ K·Hz$^{-1/2}$·W·cm$^{-2}$).[32]

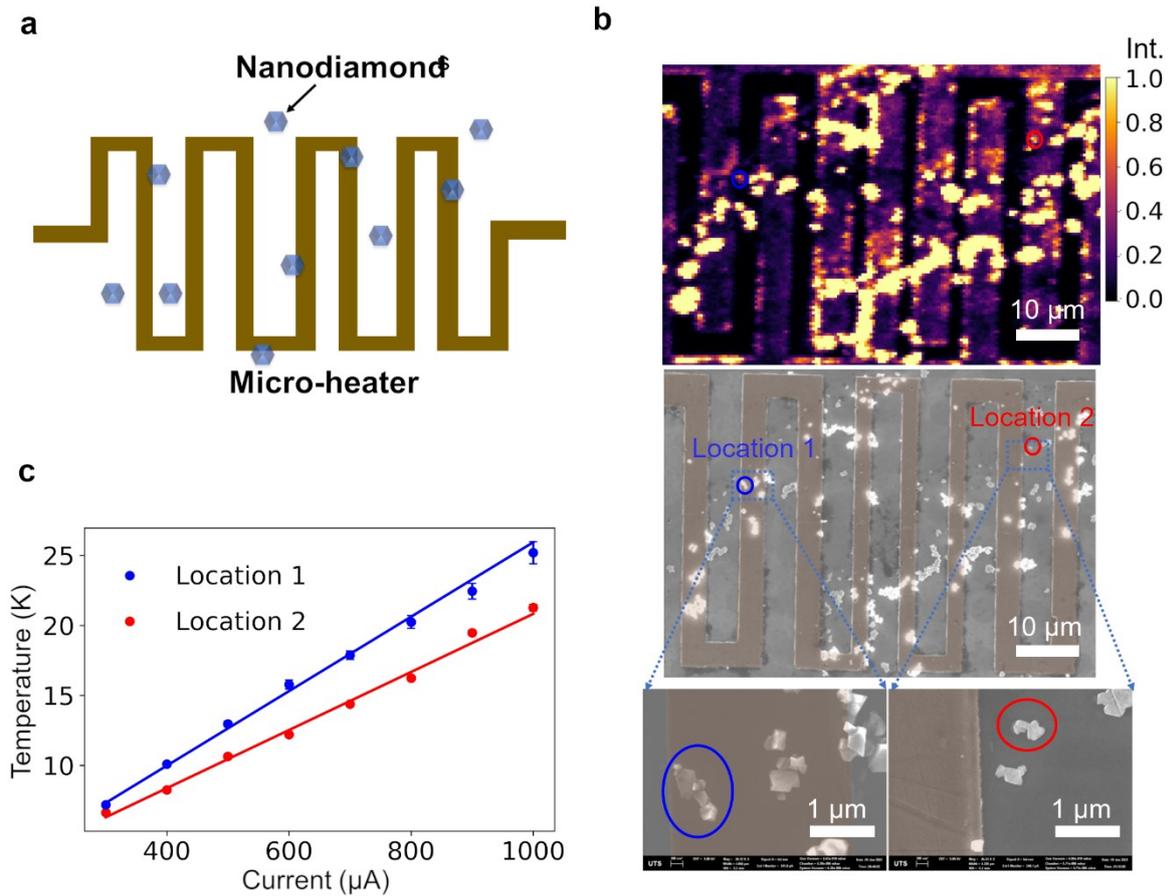

***Figure 3.*** *Temperature monitoring of an operating micro-circuit in a cryostat with a fixed temperature of 5 K. a) Schematic illustration of the microcircuit with a nanodiamond thermometer. b) Confocal map and corresponding SEM image of the microcircuit made by EBL. The confocal map is acquired with a 532-nm, 300-μW laser at 5 K. The two circled nanodiamonds are chosen to conduct temperature monitoring. c) Temperature readout of the two selected locations of the operating micro-circuit as a function of the applied current. Each temperature readout was calculated from the PLE spectra obtained with a 15-nW laser and a total scanning time of 180 s.*

Finally, we showcase the thermometric capabilities of our thermometer by monitoring the temperature in a few target locations of a microcircuit. **Figure 3a** shows a schematic of the microcircuit comprising sections of various widths: 6 μm, 4 μm and 2 μm, respectively. We fabricated the microcircuit using electron-beam lithography (EBL) and deposited a thin layer of chromium (200 nm) to form the microcircuit. The micrometer size of the circuit and its serpentine structure with different widths lead to the circuit displaying different local temperatures across it, depending on the location, when current is modulated through it. The device is therefore an ideal testbed for illustrating the performance of our nanothermometer. In the demonstration, the nanodiamonds were drop casted onto the microcircuit; the chip was heated to 60 °C for three minutes and 100 °C for a minute to completely evaporate the remaining solvent and promote strong thermal contact between the nanodiamonds and the microcircuit. To perform the temperature monitoring, we set the sample temperature to 5 K and varied the input currents to create different local temperatures on the microcircuit by Joule heating. **Figure 3b** (top panel) shows a photoluminescence (PL) confocal map that displays a number of bright spots around or on the microcircuits. A scanning electron microscopy (SEM) image taken from the same area (**Figure 3b**, middle panel) reveals that these bright spots are indeed nanodiamonds. Two selected locations of the nanodiamonds, Location 1 and Location 2, were monitored. **Figure 3b** (bottom panel) shows the close-up images of the two chosen locations in the middle panel—where two clusters of nanodiamonds with size of ~100–500 nm can be discerned. The selection was made to show two representative locations with a different heating response, as nanodiamonds closer to the microcircuit should experience higher temperature variations as current is modulated through the circuit. In each location, we performed PLE measurements on a single GeV and calibrated the center's response to temperature changes, analogously to **Figure 2a**. As we increased the input current by steps of 100 µA, we obtained the corresponding values for the linewidth $\gamma_{tot}$. We then mapped these values of the linewidth onto the calibration curves and extracted the corresponding temperatures at the two locations. **Figure 3c** shows the local temperature at the two locations plotted against the input current. We note that there seems to be a linear relationship between temperature and input current for both locations. This is likely due to the complex interplay between three factors: 1) the non-linear temperature-dependence of the resistivity of chromium [REF], 2) the heat exchange, at steady-state, between the microcircuit and the surrounding reservoir [REF], 3) Joule heating of the microcircuit [REF]. The thermometer in Location 1 experienced consistently higher temperature than that from Location 2, which is consistent with its position on the edge of the microcircuit, rather than in its close proximity. This result shows that monitoring the PLE linewidth of a single GeV can be effectively used for nanoscale thermometry at cryogenic temperatures. While our measurement was performed using GeV centers, the method can be generalized to any emitter in other host materials including diamond,[33] silicon carbide,[34] hexagonal boron nitride,[35] gallium nitride,[36] silicon nitride,[37] aluminium nitride[38] and many others.[22, 39-41] Furthermore, the current operating temperature limit of our thermometer, ~35 K, is purely a technological limitation set by the hardware of our tunable laser rather than a scientific roadblock.

## Conclusion

To conclude, we have demonstrated a cryogenic nanothermometry technique based on measuring the temperature-dependent linewidth broadening in the resonant photoluminescence spectrum of a single diamond GeV center. Our nanothermometer exhibits a record-high relative sensitivity of 20% K$^{-1}$ among all-optical nanothermometers and requires a relatively much lower laser power (~tens of nanowatts) thanks to the use of resonant excitation. To showcase the performance of our nanothermometers we employed them to accurately read out local differences in temperature from selected locations of a custom-made microcircuit. Our study demonstrates a powerful and versatile technique that allows monitoring the temperature of fundamental phenomena and/or practical devices at cryogenic conditions.

## Methods
**Sample preparation**

Ge-doped nanodiamonds were fabricated at high pressures and high temperatures in the Ge (0.2 at%) growth system. The nanodiamonds were synthesized at a pressure of 9 GPa and a temperature of 1500–1600 °C for about 60 s. For the synthesis experiments, powder mixture of Adamantane $C_{10}H_{16}$ (300 mg, > 99%, Sigma-Aldrich) and Tetraphenylgermane $C_{24}H_{20}Ge$ (15mg, 96%, Aldrich) were mixed in a mortar and pestle, both made of jasper, for about 5 min, pressed into a pellet (65 mg) and placed inside a titanium capsule (6 mm in diameter, 4 mm in height, with 0.2-mm wall thickness). A toroid-type, high-pressure chamber was used to generate pressure and temperature in the reaction cell.[42] After the treatment, samples were quenched under pressure to room temperature.

The nanodiamonds were then dispersed in isopropanol alcohol (IPA) at a concentration of 0.1% (w/w). The size of the nanodiamonds were ~300-500 nm. Five microliters of the solution were drop-cast on a clean 0.5 x 0.5 $cm^2$ silicon substrate and left to dry on a hotplate at 60 °C to completely remove the residual solvent. The silicon chip was then ready to be used for optical and structural characterization.

**Device fabrication**
Preparation of Microcircuits. A volume of ~0.1 mL of polymethyl methacrylate resist solution (950 PMMA) was spun-cast on the thermally-grown $SiO_2$ (~300 nm) on Si substrate for 1 min at 3000 rpm to get a resist coating with the thickness of ~200 nm. The resist coating was then patterned using a scanning electron microscopy (SEM) (Zeiss Supra 55VP) coupled to the Raith electron beam lithography (EBL) system. The resist pattern was subsequently formed by immersing in the resist developer (a methyl isobutyl ketone (MIBK)/isopropyl alcohol (IPA) (1:3) solution) for 30 seconds and the resist stopper (IPA) for 1 min. After cleaning the resist residual of the pattern area via $O_2$ plasma for 10 s under 50 sccm $O_2$ and 100 W power, 200 nm of chromium was deposited onto the patterned resist coating using a lab-built plasma-assisted sputter deposition chamber. The microcircuit was obtained by immersing into the resist remover (99%, acetone) to eliminate the remaining resist and excessive metal.

**Optical characterization**
The sample was mounted on the cooling stage of an Attocube Attodry800 closed-system cryostat, placed under vacuum, and cooled to 5 K. Optical measurements were then performed using a home-built scanning confocal microscope with a 0.82 NA vacuum compatible objective mounted inside the cryostat (also at 5 K). A tunable dye laser (Sirah Matisse 2 DS) with linewidth around 100 kHz was used for resonant excitation, and a 532 nm diode laser (Laser Quantum GEM) was used for off-resonant excitation. For photoluminescence excitation measurements, a scan rate of 1 GHz/s was used to tune the dye laser wavelength over the ZPL, and then the phonon sideband emission from SPE was filtered using a 600/14 nm bandpass filter, coupled via a single-mode fiber, and detected using an Avalanche Photodiode (Excelitas SPCM-AQRH). Second-order correlation measurements were performed using a fiber beam splitter and a TimeTagger20, and lifetime measurements were taken using 40 MHz pulsed 512 nm laser (PiL051X, Advanced Laser Diode Systems GmbH) and TimeTagger20.

# Notes

The authors declare no competing financial interest.

# Author contribution

T. T. T conceived the idea of the project. Y. C. made the microcircuit and samples. S. W. and Y. C. performed the optical characterization and the thermometric measurements. Y. C., S. W. and T. T. T analyzed the data. S. W. built the optical system and its software. E. E. fabricated the doped nanodiamonds. T. T. T and S.W. supervised the project. All authors discussed the results and commented on the manuscript.

# Fundings

We acknowledge the Australian Research Council (DE220100487, CE200100010, DP190101058) and the Office of Naval Research Global (N62909-22-1-2028) for financial support. The authors thank the UTS node of Optofab ANFF for the assistance with nanofabrication. E.A. Ekimov thanks the support from the Russian Science Foundation, Grant No. 19-12-00407. C Bradac thanks NSERC (RGPIN-2021-03059 and DGECR-2021-00234) and CFI JELF (#41173) for financial support.